\documentclass[aps,prd,preprint,superscriptaddress,tightenlines,floatfix,showpacs,11pt]{revtex4-1}
\usepackage{graphics}
\usepackage{graphicx}
\usepackage{amsmath,amssymb}
\usepackage{tabularx}
\usepackage[english,activeacute] {babel}

\begin{document}

\title{Accretion disk around a Schwarzschild black hole in asymptotic safety}

\author{Fabi\'an H. Zuluaga}\email{fhzuluagag@unal.edu.co}

\author{Luis A. S\'anchez}\email{lasanche@unal.edu.co}

\affiliation{Escuela de F\'\i sica, Universidad Nacional de Colombia,
A.A. 3840, Medell\'\i n, Colombia}

\begin{abstract}
We study quantum gravity effects on radiation properties of thin accretion disks around a renormalization group improved (RGI-) Schwarzschild black hole. In the infrared (IR) limit of the asymptotically safe theory with higher derivatives, the running Newton coupling $G(r)$ depends on a free parameter which encodes the quantum effects on the spacetime geometry. By varying this parameter, modifications to thermal properties of the disk as the time averaged energy flux, the disk temperature, the differential luminosity, and the conversion efficiency of accreting mass into radiation, are obtained. In addition to a shifting of the radius of the innermost stable circular orbit (ISCO) toward small values, we find an increase of the maximum values of these thermal properties and a greater efficiency than in the classical relativistic regime. We discuss astrophysical applications of these results by using observational data of the stellar-mass black hole candidate LMC X-3. Our findings could, in principle, be used to identify quantum gravity effects through astrophysical observations.\\

\noindent
{\bf keywords}: Quantum aspects of black holes, asymptotic safety, accretion disks.
\end{abstract}


\maketitle

\section{\label{sec:intr}Introduction}
Black holes are solutions to Einstein’s field equations of General Relativity (GR) describing regions of spacetime that have undergone gravitational collapse. Inside the event horizon, these solutions are plagued with singularities where the theory loses its predictive character and it is no longer possible to define a classical notion of spacetime. This forces to consider GR as an effective theory of gravity that is valid only up to certain energy scales. At high energy scales, comparable to the Planck scale, it is expected that a full theory of quantum gravity will resolve the unphysical black hole singularities so that predictivity be restored. However, attempts of describing gravity within the framework of quantum field theory face the problem of the perturbatively non-renormalizable nature of GR. As a consequence, different approaches have emerged, as loop quantum gravity and spin foams \cite{r1,r2,r3,r4}, string theory \cite{r5,r6,r7,r8}, Ho\v{r}ava-Lifshitz gravity \cite{r9,r10,r11}, causal and euclidean dynamical triangulations \cite{r12,r13}, and causal sets \cite{r14}. Another promising proposal to address this problem is the Asymptotic Safety scenario (AS) which uses the techniques of the functional renormalization group (FRG). In fact, the basic conjecture in this construction is the existence of a non-Gaussian fixed point (NGFP) of the gravitational renormalization group flow that controls the behavior of the theory at trans-Planckian energies, where the physical degrees of freedom interact predominantly antiscreening, and that renders physical quantities safe from unphysical divergences at all scales \cite{r15}. Despite AS defines a consistent and predictive quantum theory for gravity within the framework of quantum field theory, it remains a prediction since a rigorous existence proof for the NGFP is still lacking. There is, however, substantial evidence supporting the existence of the non-trivial renormalization group fixed point at the heart of this construction \cite{r16,r17}.

The importance of black holes as testing ground for gravity theories in the strong field regime has motivated numerous studies on the implications of AS gravity for black hole physics, most of them aimed at determining quantum corrections to the classical metrics, namely the RGI-Schwarzschild \cite{r18}, Kerr \cite{r19}, Schwarzschild-(A)dS \cite{r20}, Kerr-(A)dS \cite{r21}, and Reissner-Nordstr\"{o}m-(A)dS \cite{r22} spacetimes, and to study the fate of the classical singularity. Notably, it has been shown that quantum effects associated the RGI-collapse of matter lead to a softening of the singularity or even to remove it entirely \cite{r23,r24,r25,r26}. In a more phenomenological setting, quantum effects on evaporation processes and mini-black hole production in colliders has been discussed in \cite{r27,r28,r29}, corrections to the radial accretion of matter onto a RGI-Schwarzchild black hole have been analyzed in \cite{r30,r31,r32}, implications for the dark matter problem in galaxies have been studied in \cite{r33}, the implications of the gravitational antiscreening in cosmology have been considered in \cite{r34,r35,r36} (see also \cite{r37} for a recent review), and the inclusion of the matter content of the Standard Model, the so-called asymptotically safe gravity-matter models, has been explored in \cite{r38,r39,r40} (see also \cite{r41} and references therein). Nowadays, the possibility of observation of the immediate environment of black holes with angular resolution comparable to the event horizon through the Event Horizon Telescope, has provided us with the first shadow of a black hole which is visible in images of horizon-scale structures. In the AS scenario for quantum gravity, this image has opened the possibility to take a first step to bridge the gap between theory and observations by studying both the shape and size of the shadows cast by non-rotating and rotating black holes \cite{r42,r43,r44}, and the effects of simple static disk-like structures on these shadows \cite{r45}. 

Even though the metric is strongly modified close the classical singularity, quantum effects persist even outside the horizon. This has been shown by explicit calculation of the effects on the dynamics of test particles around RGI-Schwarzchild and Kerr black holes in AS with higher derivatives in the IR-limit \cite{r46}. Since the IR-regime is obtained in the limit $r \gg l_{\rm Planck}$ and since the description of the motion of test particles around black holes is in the basis of the construction of realistic models of thin accretion disks, its is clear that effects on the physical properties of these disks should be present. The study of such effects can lead to the identification of specific signatures that allow to distinguish the RGI-geometries by using astrophysical observations. With this purpose in mind, X-ray reflection spectroscopy of a Novikov-Page-Thorne type disk \cite{r47,r48} has been used in \cite{r49} to constraint the dimensionless fixed-point parameter appearing in the rotating black hole RGI-metric proposed in \cite{r42}. In the same token, the iron line shape expected in the reflection spectrum of accretion disks around black holes in the IR-limit of the AS gravity with higher derivatives has been studied in \cite{r50}.

As an additional effort to paving the way to confront the AS theory with observations, in this paper we study quantum gravity effects on radiation properties of a thin accretion disk, the relativistic Novikov-Page-Thorne model, surrounding a RGI-Schwarzschild black hole in the context of the IR-limit of the AS theory with higher derivatives. In particular, the time averaged energy flux, the disk temperature, the differential luminosity, and the conversion efficiency of accreting mass into radiation, are obtained and compared with the classical solution.

The outline of this paper is as follows. In Sec.~\ref{sec:sec2} we present the RGI-Schwarzschild metric in the AS theory with higher derivatives. In Sec.~\ref{sec:sec3} the equations describing the equatorial geodesics for massive particles are obtained. In Sec.~\ref{sec:sec4} we discuss the standard thin accretion disk model and present the equations describing its thermal properties. In Sec.~\ref{sec:sec5} we calculate the properties of thin accretion disks in the RGI-geometry, analyze the quantum gravity effects by comparing with the GR description, and discuss astrophysical applications of our results by using observational data of the black hole candidate LMC X-3. Finally, in the last section, we present our conclusions.
%
\section{\label{sec:sec2}Renormalization group improved Schwarzschild metric}
The study of black hole solutions in asymptotic safety with higher derivative terms, in which the effective action is not restricted to the Einstein-Hilbert term, but also includes the Ricci scalar square, the Ricci tensor square, the Kretschmann scalar and running gravitational couplings, has been done in \cite{r51,r52} and has been further analyzed in \cite{r53,r54,r55}. In the infrared (IR) limit, the quantum gravity corrections to the classical Schwarzschild metric are accounted for by promoting, as usual, the Newton coupling $G_0$ through a running coupling $G(r)$ \cite{r52}:
\begin{equation}\label{eq1}
G_0 \rightarrow G(r) = G_0\left(1-\frac{\xi}{r^2}\right),
\end{equation}
where $\xi$ is a parameter with dimensions of length squared associated to the scale identification between the momentum scale $p$ and the radial coordinate $r$ which, in the IR limit, takes the form $p \sim 1/r$. Thus, the line element of the RGI-Schwarzschild metric for a black hole with mass $M$ remains spherically symmetric and takes the form
\begin{equation}\label{eq2}
ds^2=-f(r)dt^2+f(r)^{-1}dr^2+r^2\Big(d\theta^2+\sin^2\theta d\phi^2\Big),
\end{equation}
with the metric coefficient $g_{tt}$, in units $c =G_{0}= 1$, given by
\begin{equation}\label{eq3}
f(r) = 1-\frac{2M}{r}\left(1-\frac{\xi}{r^2}\right).
\end{equation}
The radius of the new event horizon comes from $f(r)=0$, that is
\begin{equation}\label{eq4}
r^3-2 M r^2+2 M^3 \tilde{\xi} = 0,  
\end{equation}
where the dimensionless parameter $\tilde{\xi}=\xi/M^2$ has been introduced for convenience. The only real solution $r_{AS}$ to Eq.(\ref{eq4}) is
\begin{eqnarray} \nonumber
\frac{r_{AS}}{r_{S}}&=&\frac{1}{3}+\frac{2}{3\sqrt[3]{8-27 \tilde{\xi}+3 \sqrt{3} \sqrt{\tilde{\xi} (27 \tilde{\xi} -16)}}} \\ \label{eq5}
&& +\frac{1}{6}\sqrt[3]{8-27 \tilde{\xi}+3 \sqrt{3} \sqrt{\tilde{\xi} (27 \tilde{\xi} -16)}}, 
\end{eqnarray}
where $r_{S}=2M$ is the classical Schwarzschild radius. Note that for $\tilde{\xi}=0$ we have $r_{AS}=r_{S}$ as expected, while for $\tilde{\xi}$ greater than the critical value $\tilde{\xi}_{c}=16/27$, there is no horizon at all and a naked singularity arises. This means that each value of $\tilde{\xi}$ in the range $0 \leq \tilde{\xi} \leq \tilde{\xi}_{c}$ picks out a critical mass $M_{c}=\sqrt{27\xi/16}$ in such a way that for $M>M_{c}$ there are two horizons: one inner Cauchy horizon and one outer event horizon, for $M=M_{c}$ the two horizons merge, and for $M<M_{c}$ a naked singularity develops. Since our aim is to discuss quantum gravity effects on the radiation properties of accretion disks around a Schwarzschild black hole, we will not be concerned with naked space-time singularities in this work.

\begin{figure}[!ht]
\centering
\includegraphics[scale=1.0]{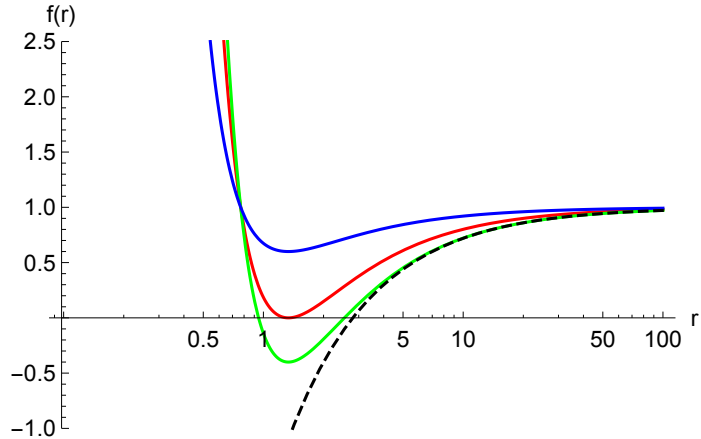}
\caption{\label{Fig1} 
Plots of the improved metric coefficient $f(r)$ for $M=0.4<M_{c}$ (blue), $M=M_{c}$ (red), and $M=1.4>M_{c}$ (green), with $M_{c}=1$ corresponding to $\xi = 16/27$. The dashed line shows the classical $f_0(r)$ for $M=1.4$}
\end{figure}

\noindent
In Fig.~\ref{Fig1} we show plots of $f(r)$ vs. $r$ for $M<M_{c}$ (blue), $M=M_{c}$ (red), and $M>M_{c}$ (green), with $M_{c}=1$ corresponding to $\xi = 16/27$. For the shake of comparison, the dashed line shows the classical metric coefficient $f_0(r)=1-2Mr^{-1}$ for $M=1.4$. Clearly, the quantum effect amounts to a shifting of the horizon of the RGI-black hole solution toward smaller values with respect to its classical counterpart.

%
\section{\label{sec:sec3}Geodesics in the RG-improved geometry}
The line element in Eq.~(\ref{eq2}) can be written in the form of a Schwarzschild-like metric by introducing an ``effective'' mass $M_{\rm eff}(r)$ which is function of the radial coordinate and depends on the free parameter $\tilde{\xi}$, that is
\begin{eqnarray}\nonumber
ds^2&=&-\left(1-2\frac{M_{\rm eff}(r)}{r}\right)dt^2+\left(1-2\frac{M_{\rm eff}(r)}{r}\right)^{-1}dr^2 \\ \label{eq6}
&& +r^2\Big(d\theta^2+\sin^2\theta d\phi^2\Big),
\end{eqnarray}
with
\begin{equation}\label{eq7}
M_{\rm eff}(r)=M\left(1-\frac{\xi}{r^2}\right).
\end{equation}
Using this definition is immediate to write the Lagrangian for the improved metric by doing $M \rightarrow M_{\rm eff}(r)$ in the classical expression. Then, restricting ourselves to the equatorial plane ($\theta=\pi/2$, $\dot{\theta}=0$), the Lagrangian is written as
\begin{eqnarray}\nonumber
2L&=&-\left(1-2\frac{M_{\rm eff}(r)}{r}\right)\dot{t}^2+\left(1-2\frac{M_{\rm eff}(r)}{r}\right)^{-1}\dot{r}^2 \\ \label{eq8}
&& +r^2 \dot{\phi}^2, 
\end{eqnarray}
where the dots denote derivative with respect to an affine parameter that, for massive particles which follows timelike geodesics, can be taken as the proper time $\tau$.
The generalized momenta are
\begin{eqnarray}\label{eq9}
p_t&=&-\left(1-2\frac{M_{\rm eff}(r)}{r}\right)\dot{t}=-k, \\ \label{eq10}
p_r&=&\left(1-2\frac{M_{\rm eff}(r)}{r}\right)^{-1}\dot{r}, \\ \label{eq11}
p_{\phi}&=&r^2 \dot{\phi}=h. 
\end{eqnarray}
Here the constants $k$ and $h$ are, respectively, the energy and angular momentum per unit rest mass of the particle describing the trajectory.
The Hamiltonian $H=p_t\dot{t}+p_r\dot{r}+p_{\phi}\dot{\phi}-L$, which is constant as it is $t$ independent, is then given by 
\begin{equation}\label{eq12}
2H=-k \dot{t}+\left(1-2\frac{M_{\rm eff}(r)}{r}\right)^{-1}\dot{r}^2+h\dot{\phi}=-1,
\end{equation}
where the last equality follows from the fact that the particle is assumed at rest at infinity.

\noindent
Solving Eqs.~(\ref{eq9}) and (\ref{eq11}) for $\dot{t}$ and $\dot{\phi}$ and substituting into Eq.~(\ref{eq12}) we obtain the ``energy'' equation
\begin{equation}\label{eq13}
\frac{1}{2}\dot{r}^2+V_{\rm eff}(r)=\frac{1}{2}(k^2-1),
\end{equation}
where the effective potential per unit mass is given by
\begin{equation}\label{eq14}
V_{\rm eff}(r)=-\frac{M_{\rm eff}(r)}{r}+\frac{h^2}{2r^2}-\frac{M_{\rm eff}(r)h^2}{r^3}.
\end{equation}
Another useful equation to describe the orbits of massive particles gives us $r$ as a function of $\phi$ (see, e.g., \cite{r56}). It can be obtained by using Eq.~(\ref{eq11}) to write
\begin{equation}\label{eq15}
\dot{r}=\frac{dr}{d\tau}=\frac{dr}{d\phi}\frac{d\phi}{d\tau}=\frac{h}{r^2}\frac{dr}{d\phi}.
\end{equation}
Substituting into Eq.~(\ref{eq13}) and making the change of variable $u=1/r$, we have
\begin{equation}\label{eq16}
\left(\frac{du}{d\phi}\right)^2+u^2=\frac{k^2-1}{h^2}+\frac{2uM_{\rm eff}(u)}{h^2}+2 u^3 M_{\rm eff}(u).
\end{equation}
Differentiating this equation with respect to $\phi$ we finally obtain
\begin{equation}\label{eq17}
\frac{d^2u}{d\phi^2}+u=\frac{M}{h^2}(1-3\xi u^2)+M u^2 (3-5\xi u^2).
\end{equation}

\subsection{\label{sec:sec3.1}Circular orbits of massive particles}
To determine the basic equations describing the time averaged radial disk structure, we first explicitly calculate the specific angular momentum $h$, the specific energy $k$ and the angular velocity $\Omega$ of particles moving in circular trajectories.

For circular orbits in the equatorial plane $\dot{r}=0$ such that $r$ (and $u=1/r$) is constant. Thus, from Eq.~(\ref{eq17}) we obtain an expression for the specific angular momentum $h$ that we write, for convenience, in terms of the dimensionless quantities $x=r/M$ and $\tilde{\xi}=\xi/M^2$:
\begin{equation}\label{eq18}
h=\frac{M x \sqrt{x^2-3 \tilde{\xi}}}{\sqrt{x^3-3x^2+5\tilde{\xi}}}.
\end{equation}
Making $\dot{r}=0$ in Eq.~(\ref{eq13}) and substituting the expression for $h$ allows us to identify the specific energy $k$:
\begin{equation}\label{eq19}
k=\frac{x^3-2 x^2+2\tilde{\xi}}{x^{3/2}\sqrt{x^3-3x^2+5\tilde{\xi}}}.
\end{equation}
Using Eqs.~(\ref{eq18}) and (\ref{eq19}) the angular velocity can be calculate from Eqs.~(\ref{eq9}) and (\ref{eq11}) as
\begin{equation}\label{eq20}
\Omega=\frac{\dot{\phi}}{\dot{t}}=\frac{1}{M x^{5/2}}\sqrt{x^2-3\tilde{\xi}}.
\end{equation}
Upon substituting Eq.~(\ref{eq18}) into Eq.~(\ref{eq14}), the effective potential acquires the form
\begin{eqnarray} \nonumber
V_{\rm eff} =&-&\frac{x^2-\tilde{\xi}}{x^3}\left(1+\frac{x^2-3\tilde{\xi}}{x^3-3x^2+5\tilde{\xi}} \right)\\ \label{eq21}
&+&\frac{1}{2}\frac{x^2-3\tilde{\xi}}{x^3-3 x^2+5\tilde{\xi}}.
\end{eqnarray}
It must be noticed that $V_{\rm eff}$, $h$, $k$ and $\Omega$ reduce to the classical expressions in the limit $\tilde{\xi} \rightarrow 0$.

\noindent
Circular orbits occur at the local minimum of the effective potential. Thus, the dimensionless radius of the innermost stable circular geodesic orbit $x_{\rm isco}$ is calculated from
\begin{equation}\label{eq22}
\frac{d^2V_{\rm eff}}{dx^2}=0,
\end{equation}
where, due to the fact that $h$ is constant for circular orbits, the derivatives mus be calculated for $V_{\rm eff}$ as given by Eq.~(\ref{eq14}) \cite{r57,r58,r59,r60}. This yields
\begin{equation}\label{eq23}
\frac{d^2V_{\rm eff}}{dx^2}=-\frac{2x^2(x^2-6\tilde{\xi})-3\bar{h}^2(x^2(x-4)+10\tilde{\xi})}{x^7},
\end{equation}
where $\bar{h}=h/M$. Substituting Eq.~(\ref{eq18}) into (\ref{eq23}) we have
\begin{equation}\label{eq24}
\frac{d^2V_{\rm eff}}{dx^2}=\frac{x^4(x-6)+x^2(3x+20)\tilde{\xi}-30\tilde{\xi}^2}{(x-3)x^7+5x^5\tilde{\xi}}.
\end{equation}
\begin{figure}[b]
\centering
\includegraphics[scale=1.0]{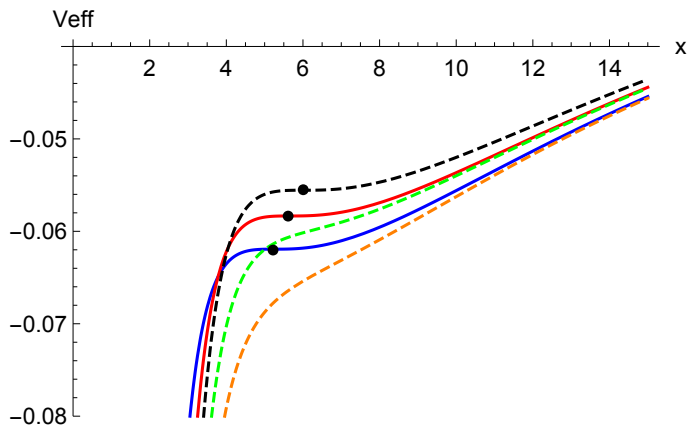}
\caption{\label{Fig2} 
The effective potential for the values of $\bar{h}=h/M$ when evaluated at the ISCO and for $\tilde{\xi}=\tilde{\xi}_c=16/27$ (solid blue curve), $\tilde{\xi}=0.3$ (solid red curve) and for the classical case $\tilde{\xi}=0$ (black dashed curve). The green and orange dashed lines shows the GR effective potential for the values of $\bar{h}$ corresponding to the RGI-black hole for the same values of $\tilde{\xi}$. The dots indicate the locations of the ISCO for each of those cases.}
\end{figure}
\begin{figure}[t]
\centering
\includegraphics[width=8.0cm,height=5.5cm]{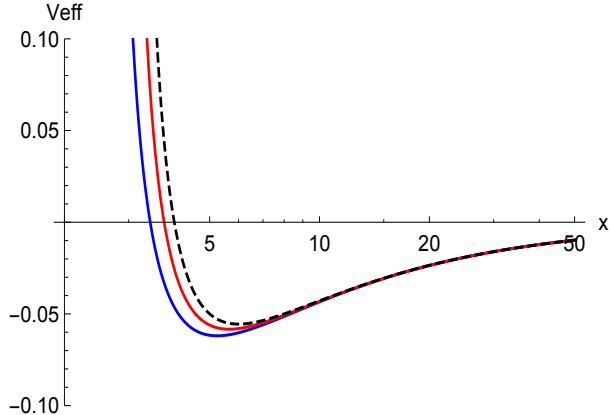}
\caption{\label{Fig3}
The effective potential $V_{\rm eff}$ vs. $x$ for several values of $\tilde{\xi}=\xi/M^2$. From left to right: $\tilde{\xi}=\tilde{\xi}_c=16/27$ (blue), $\tilde{\xi}=0.3$ (red). The classical case $\tilde{\xi}=0$ is given by the black dashed curve} 
\end{figure}
In Table \ref{tab:1} we report the values of the dimensionless inner boundary of the disk for selected values of $\tilde{\xi}$.
\begin{table}[!ht]
\caption{\label{tab:1}The ISCO of the thin disk for selected values of $\tilde{\xi}$}
 \begin{ruledtabular}
 \begin{tabular}{ll}
 $\tilde{\xi}$ & $x_{\rm isco}$ \\ \hline
 16/27 & 5.2439 \\
 0.30 & 5.6559 \\
 0 & 6 \\
 \end{tabular}
 \end{ruledtabular}
\end{table}
\noindent
Alternatively, $x_{\rm isco}$ can be computed from \cite{r48}
\begin{equation}\label{eq25}
\frac{dh}{dx}=\frac{dk}{dx}=0.
\end{equation}

\begin{figure}[b]
\centering
\includegraphics[scale=1.0]{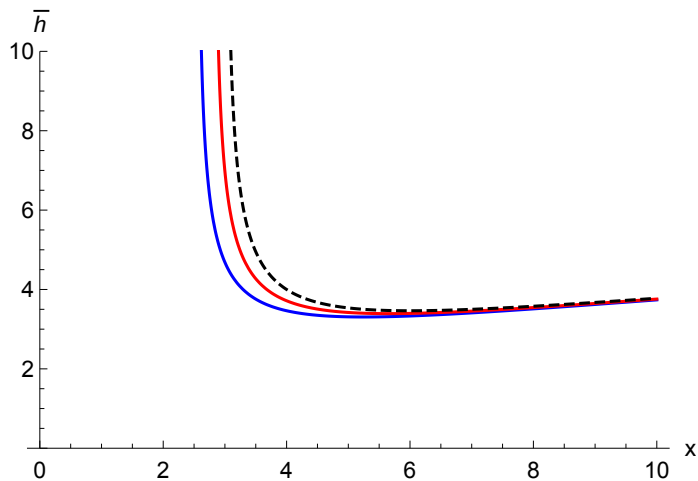}
\caption{\label{Fig4}
The angular momentum $\bar{h}=h/M$ vs. $x$ for several values of $\tilde{\xi}=\xi/M^2$. From left to right: $\tilde{\xi}=\tilde{\xi}_c=16/27$ (blue), $\tilde{\xi}=0.3$ (red). The dashed curve corresponds to the classical case $\tilde{\xi}=0$} 
\end{figure}
\begin{figure}[t]
\centering
\includegraphics[scale=1.0]{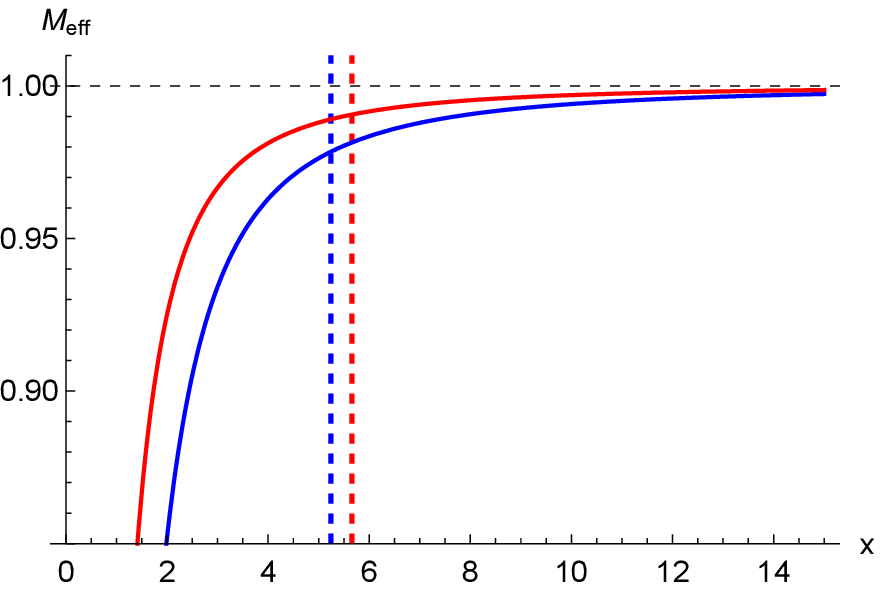}
\caption{\label{Fig5}
Plots of the effective mass as a function of $x$ for $\tilde{\xi}=\tilde{\xi}_c=16/27$ (blue) and $\tilde{\xi}=0.3$ (red). The vertical dashed lines indicate the ISCO for the same values of $\tilde{\xi}$, respectively.}
\end{figure}
Fig.~\ref{Fig2} shows the effective potential (cf. Eq.(\ref{eq14})) for the values of $\bar{h}=h/M$ when evaluated at the ISCO and for $\tilde{\xi}=\tilde{\xi}_c=16/27$ (solid blue), $\tilde{\xi}=0.3$ (solid red) and for the classical case $\tilde{\xi}=0$ (black dashed line). The green and orange dashed lines show the GR effective potential for the values of $\bar{h}$ corresponding to the RGI-black hole when $\tilde{\xi}=\tilde{\xi}_c=16/27$ and $\tilde{\xi}=0.3$, respectively. The dots indicate the locations of the ISCO for each of these cases. Its is evident that for $\tilde{\xi}\neq 0$ only the effective potential for the RGI-Schwarzschild black hole develops a minimum which shifts to smaller values of $x$. In Figs.~\ref{Fig3} and \ref{Fig4} we plot $V_{\rm eff}$ (cf. Eq.(\ref{eq21})) and $\bar{h}=h/M$ respectively, as functions of $x$ for the same values of the free parameter $\tilde{\xi}$ in the range $0\leq \tilde{\xi}\leq 16/27$, and compare with the classical solution. From these figures and from Table \ref{tab:1} it also follows that quantum gravity effects shift the radius of the innermost stable circular orbit to a smaller value and, at the same time, shift the minimal values of the effective potential and the angular momentum ${\bar h}$ toward slightly smaller values in the vicinity of $x_{\rm isco}$. It can be easily verified that a similar behavior is found in a plot of $k$ vs. $x$.

\noindent
Eqs.(\ref{eq14}) and (\ref{eq18}) show that two quantum gravity effects combine to deepening the potential well and shifting $x_{\rm isco}$ inward: (i) the presence of $M_{\rm eff} < M$ and, (ii) the decrease of the values of $h$ for $\tilde{\xi} \neq 0$. Fig.~\ref{Fig5} is a plot of the effective mass as a function of $x$ which illustrates the fact that, for a fixed value of $\bar{\xi}$, a particle falling into the black hole ``see'' a mass $M_{\rm eff}$ that decreases for decreasing values of $x$. Thus, even though the angular momentum also decreases, as the particle spirals into the black hole it feels a progressively weaker gravitational pull which allows it to remain in a stable circular orbit with smaller $x_{\rm isco}$. 

\section{\label{sec:sec4}Relativistic thin accretion disk}
The simplest non-relativistic model of an accretion disk around a compact central object assumes that matter spirals inwards losing angular momentum which, because of turbulent viscosity, is transferred outward through the disk. As the gas moves inwards, loses gravitational energy and heat up emitting thermal energy \cite{r61}. A general relativistic treatment of an accretion disk around a black hole was for the first time presented in \cite{r47,r48}. This model assumes a disk in a quasi-steady state lying in the equatorial plane of an stationary, axially-symmetric background space-time geometry, with the disk material moving in nearly geodesic circular orbits. The disk is thin that is, its maximum half-thickness $H$ is such that $H/R << 1$ with $R$ being the characteristic radius of the disk. The heat generated by stress and dynamical friction is efficiently emitted in the form of radiation mainly from the disk surface, and the quantities describing the thermal properties of the disk are averaged over the azimuthal angle $\phi=2\pi$, over the height $H$, and over the time scale $\Delta t$ that the gas takes to flow inward a distance $2H$. With these assumptions, the time-averaged radial structure of the disk is obtained from the laws of conservation of rest mass, energy and angular momentum. From the integration of the equation of mass conservation it follows the constancy of the mass accretion rate
\begin{equation}\label{eq26}
\dot{M}= - 2 \pi r\Sigma (r) u^r = {\rm constant},
\end{equation}
where $\Sigma (r)$ is the surface density of the disk and $u^r$ is the radial velocity.
\noindent
The combination of the laws of energy and angular momentum conservation provides us with the differential of the luminosity ${\cal L_{\infty}}$ at infinity \cite{r48,r62}
\begin{equation}\label{eq27}
\frac{d{\cal L_{\infty}}}{d{\rm ln}r}=4 \pi r\sqrt{-g}k {\cal F}(r),
\end{equation}
where the flux of radiant energy ${\cal F}$ emitted from the upper face of disk in the local frame of the accreting fluid is expressed in terms of the specific angular momentum $h$, the specific energy $k$ and the angular velocity $\Omega$ by
\begin{eqnarray}\nonumber
{\cal F}(r)=&-&\frac{\dot{M}}{4\pi\sqrt{-g}}\frac{1}{(k-\Omega h)^2}\frac{d\Omega}{dr} \\ \label{eq28}
& & \times \int_{r_{\rm isco}}^{r}(k-\Omega h) \frac{dh}{dr}dr,
\end{eqnarray}
where $\sqrt{-g}=r$ both for the RGI-metric and for the classical Schwarzschild spacetime. The numerical integration of 
Eq.~(\ref{eq28}) is eased by using the relation $dk/dr=\Omega (dh/dr)$ \cite{r48} and integrating by parts:
\begin{eqnarray}\nonumber
\int_{r_{\rm isco}}^{r}&(k-\Omega h)&\frac{dh}{dr}dr \\ \label{eq29}
& & =kh - k_{\rm isco}h_{\rm isco} - 2 \int_{r_{\rm isco}}^{r} h \frac{dk}{dr}dr.
\end{eqnarray} 
\noindent
Since the disk is assumed in thermodynamic equilibrium, the radiation emitted can be considered as a black body radiation with the temperature given by
\begin{equation}\label{eq30}
T(r)=\sigma^{-1/4} {\cal F}(r)^{1/4},
\end{equation}
where $\sigma$ is Stefan-Boltzmann constant.

\noindent
Finally, provided that all the emitted photons can escape to infinity, the conversion efficiency of accreting mass into radiation $\epsilon$ is obtained from the energy loss by a test particle moving from infinity to the inner boundary of the disk. Then, considering that for $r\rightarrow \infty$, $k_{\infty}\approx 1$, we have:
\begin{equation}\label{eq31}
\epsilon=\frac{k_{\infty}-k_{\rm isco}}{k_{\infty}}\approx 1 - k_{\rm isco}.
\end{equation}

\section{\label{sec:sec5}Radiation from thin accretion disks in AS with higher derivatives}
Let us now discuss the radiation properties of thin accretion disks around a RGI-Schwarschild black hole in AS gravity with higher derivatives and contrast with the predictions of general relativity. We assume the mass $M$ of the black hole an the mass accretion rate $\dot{M}$ as constant values provided the observations \cite{r50,r51,r52,r53}, thus, for comparison purposes, in what follows we take $M=1$ and we compute the thermal properties of the disk in units of mass accretion rate \cite{r56}.

\begin{figure}[b]
\centering
\includegraphics[scale=1.0]{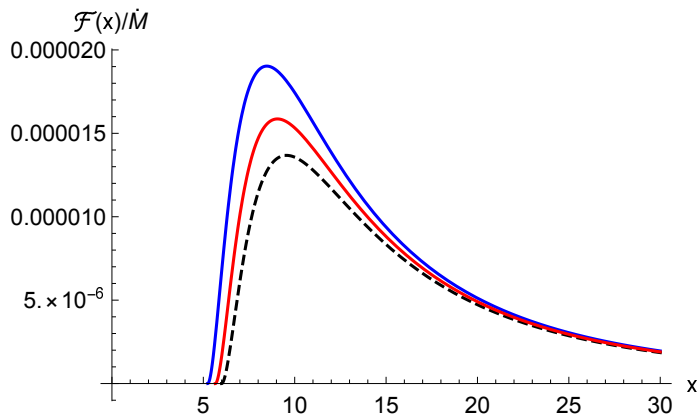}
\caption{\label{Fig6}
Energy flux per unit accretion rate from a thin accretion disk around a RGI-Schwarzschild black hole for $\tilde{\xi}=\tilde{\xi}_c=16/27$ (blue) and $\tilde{\xi}=0.3$ (red). The black dashed curve is the energy flux from the disk around a classical Schwarzschild black hole ($\tilde{\xi}=0$)} 
\end{figure}

\begin{figure}[b]
\centering
\includegraphics[scale=1.0]{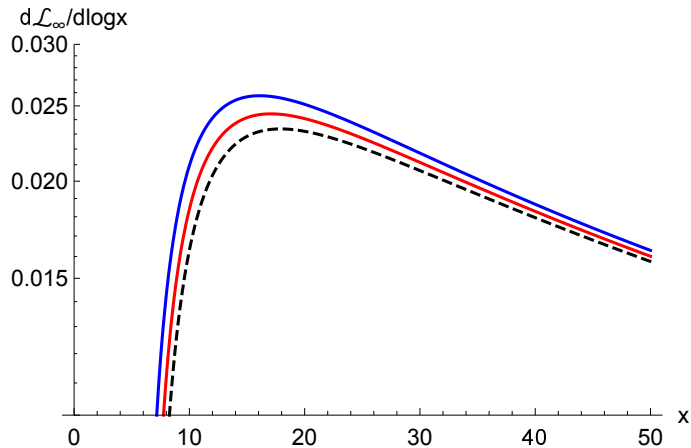}
\caption{\label{Fig7}
Plots of the differential luminosity at infinity per unit accretion rate from a thin disk around a RGI-Schwarzschild black hole for $\tilde{\xi}=\tilde{\xi}_c=16/27$ (blue) and $\tilde{\xi}=0.3$ (red). The dashed line shows the same for a disk around the classical solution ($\tilde{\xi}=0$)} 
\end{figure}

\begin{figure}[t]
\centering
\includegraphics[scale=1.0]{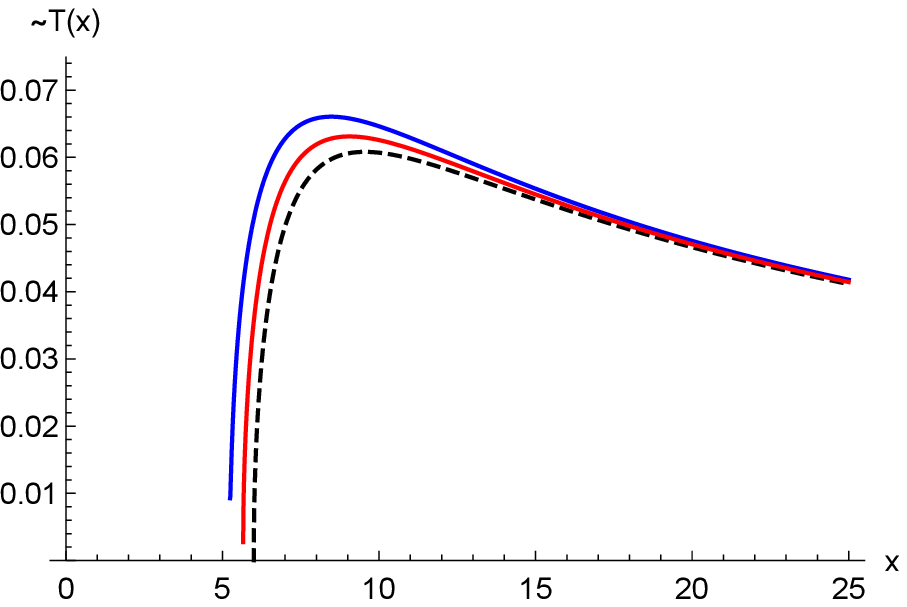}
\caption{\label{Fig8}
Radial profiles of the temperature per unit accretion rate of a thin accretion disk around a RGI-Schwarzschild black hole for $\tilde{\xi}=\tilde{\xi}_c=16/27$ (blue) and $\tilde{\xi}=0.3$ (red). The dashed line corresponds to the temperature of the disk around a Schwarzschild black hole in GR ($\tilde{\xi}=0$)} 
\end{figure}

\noindent
In Figs.~\ref{Fig6}, \ref{Fig7} and \ref{Fig8}, we plot the radial profiles of the time averaged energy flux per unit accretion rate, the differential luminosity per unit accretion rate, and the radial profile of the temperature of the accretion disk (more precisely, the radial profile of $\sigma^{1/4}T/\dot{M}^{1/4}$), respectively, for different values of the dimensionless free parameter $\tilde{\xi}$ namely, for the critical value $\tilde{\xi}_c=16/27$, for $\tilde{\xi}=0.3$ and for the classical solution $\tilde{\xi}=0$. From Fig.~\ref{Fig6} we see that with the increase of $\tilde{\xi}$ more energy is radiated away from the disk around the RGI-black hole than from the classical one, and that, associated to the shifting of the inner edge of the disk toward smaller values of $x$, there is both an increase and an inward shifting of the peak of the radial profile of the flux. In particular, for $\tilde{\xi}_c=16/27$ the increase of the maximum of the profile is $\sim 39\%$, while for $\tilde{\xi}=0.3$ the percentage change is $\sim 17\%$. Figs.~\ref{Fig7} and \ref{Fig8} show that these effects are also present in the differential luminosity and in the temperature of the accretion disk. For $\tilde{\xi}=\tilde{\xi}_c=16/27$, the peak of the differential luminosity is $\sim 10\%$ higher than in the general relativistic case, and the disk is $\sim 8\%$ hotter when it surrounds the RGI-black hole.


In table \ref{tab:2} we record the values of the specific energy at the ISCO $k_{\rm isco}$ and the corresponding values of the efficiency $\epsilon$ for conversion of accreted mass into radiation, when the photon capture by the black hole is neglected, for the same values of $\tilde{\xi}$ mentioned above. We see that the value of $\epsilon$ always increases with the increase of $\tilde{\xi}$ which means that the RGI-Schwarzschild black hole is a more efficient engine for conversion of accreted mass into radiation. 
\begin{table}[!ht]
 \caption{\label{tab:2}The specific energy at the ISCO and the efficiency $\epsilon$ of the conversion of the accreted mass into radiation for several values of $\tilde{\xi}$}
 \begin{ruledtabular}
 \begin{tabular}{lll}
 $\tilde{\xi}$ & $k_{\rm isco}$ & $\epsilon$ (\%) \\ \hline
 16/27 & 0.9360 & 6.40 \\
 0.30 & 0.9398 & 6.02 \\
 0 & 0.9428 & 5.72 \\
 \end{tabular}
 \end{ruledtabular}
 \end{table}
\noindent

The increase of the peak of the thermal properties of the disk and of the efficiency for conversion of accreted mass into radiation is directly linked to the deeper potential well associated to the RGI-black hole and to the shifting of the ISCO to smaller values. In fact, a smaller ISCO radius allows gravitational energy to be extracted from deeper into the steep potential well such that the disc material can convert more gravitational energy into thermal energy. 

Since a smaller ISCO can be associated to a non-zero black hole spin, the question arises about the possible astrophysical applications of our results. This, in turn, requires answering the question whether values of $\tilde{\xi}$ in the allowed range $0\leq \tilde{\xi}\leq 16/27$ can mimic the spin of a Kerr black hole in the sense that the spin parameter $a_{\ast}=a/M$ and the dimensionless parameter $\tilde{\xi}$ give rise to the same ISCO radius. To address this issue we recall that the ISCO radius for co-rotating circular orbits around a Kerr black hole is given by \cite{r63}
\begin{equation}\label{eq32}
r_{\rm isco}=3 + Z_2 - \sqrt{(3-Z_1)(3+Z_1+2Z_2)},  
\end{equation}
where
\begin{eqnarray}\nonumber
Z_1 &=&1 + \sqrt[3]{1-a^2}(\sqrt[3]{1-a}+\sqrt[3]{1+a}), \\ \nonumber
Z_2 &=& \sqrt{3 a^2 + Z_1^2} \nonumber
\end{eqnarray}
The ISCO of the thin accretion disk around the RGI-Schwarzschild black hole comes from solving Eq.(\ref{eq22}) using Eq.(\ref{eq24}), that is
\begin{equation}\label{eq33}
x^4(x-6)+x^2(3x+20)\tilde{\xi}-30\tilde{\xi}^2=0.  
\end{equation}
%
\begin{figure}[t]
\centering
\includegraphics[scale=1.0]{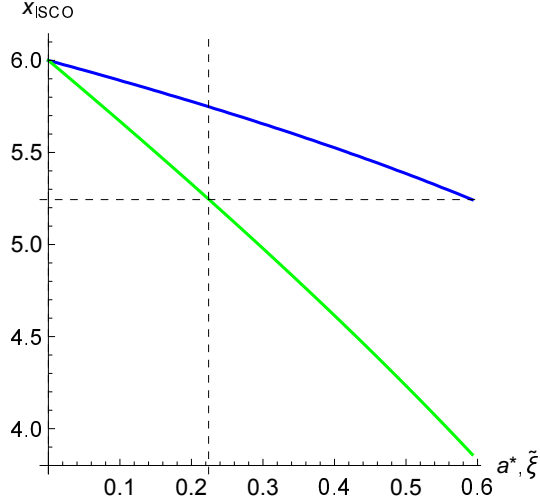}
\caption{\label{Fig9}
Dependence of the ISCO radius on the parameter $a_{\ast}$ for a Kerr black hole (green curve) and on the parameter $\tilde{\xi}$ for the RGI-Schwarzschild black hole (blue curve).} 
\end{figure}
Fig.~\ref{Fig9} illustrates the relation between the parameters $a_{\ast}$ and $\tilde{\xi}$ where the intersection of the vertical dashed line with the horizontal axis allows us to conclude that the parameter $\tilde{\xi}$ can mimic $a_{\ast}$ up to a maximum value $a^{\rm AS}_{\ast}=0.2241$ which is obtained when the dimensionless parameter $\tilde{\xi}$ takes its critical value $\tilde{\xi}_c=16/27$. This means that the the RGI-Schwarzschild black hole we are studying can mimic very slowly rotating black holes. An interesting example is the extragalactic stellar-mass black hole LMC X-3 which is a X-ray binary in the class of Roche-lobe overflowing, transient black-hole systems with $a_{\ast}=0.21^{+0.18}_{-0.22}$ (90\% C.L.) \cite{r64}, with mass $M=6.98\pm 0.56 M_{\odot}$ \cite{r65}, and that features a fixed inner-disk radius associated with the general relativistic ISCO \cite{r66}. These values do not differ much from those reported recently in \cite{r66} where $a_{\ast}$ is determined in the range $0.22-0.41$ (90\% C.L.) and $M$ is determined in the range $5.35-6.22 M_{\odot}$. Thus, assuming the relativistic thin accretion disk model as appropriate for describing the disk around LMC X-3 (see \cite{r67,r68} for a discussion on limitations of this assumption) and since $a^{\rm AS}_{\ast}$ is within the uncertainty range of the value obtained from observations, we can fix the dimensionless parameter $\tilde{\xi}$ to the value $\tilde{\xi}\simeq 16/27=0.5926$. Reinstating units and writing $G_0=M^{-2}_{\rm Pl}$, the value of $\tilde{\xi}$ translates into the following value for $\xi$
\begin{equation}\label{eq34}
\xi=\tilde{\xi}\frac{M^2}{M^{4}_{\rm Pl}}\simeq 4.41\times 10^{92},  
\end{equation}
for $M \approx 6.5 M_{\odot}$. As pointed out in \cite{r42}, such a huge number is obviously associated to the fact that AS gravity fixes the scale of new physics to be the Planck scale.

It is important to notice that the results obtained in this section show that quantum gravity effects from the AS theory with high-derivative terms are not restricted to the interior of the horizon of a black hole or its immediate vicinity, but persist beyond the horizon, even giving rise to modifications of the thermal properties of the accretion disk surrounding a Schwarzschild black hole. This should motivate the study of quantum gravity effects on more realistic black hole environments, such as accretion disks around a RGI-Kerr black hole.

\section{\label{sec:sec6}Conclusions and final remarks}
In this work we have studied, for the first time at the best of our knowledge, quantum gravity corrections to the thermal properties of relativistic thin accretion disk around a RGI-Schwarzschild black hole in the IR-limit of the asymptotic safety scenario for quantum gravity with high-derivative terms. We have calculated, in particular, corrections to the time averaged energy flux, the differential luminosity at infinity, the disk temperature and the conversion efficiency of accreting mass into radiation. When compared to the predictions of general relativity, we have found that an increase of the parameter $\tilde{\xi}$, which encodes the quantum effects on the space-time geometry, not only leads to a shifting of the radius of the inner edge of the disk, the ISCO, toward smaller values, but, as a consequence, we also found: (i) an increase of the energy radiated away from the disk, (ii) an increase of the temperature of the disk, (iii) an increase of the differential luminosity reaching an observer at infinity, (iv) a shifting of the peak of the radial profiles of these thermal properties toward smaller values of the radial coordinate, and (v) a greater conversion efficiency of accreting mass into radiation. Our results show that the RGI-Schwarzschild black hole we are considering, can mimic the spin parameter of a Kerr black hole up to $a^{\rm AS}_{\ast}=0.2241$, a value obtained when the dimensionless parameter $\tilde{\xi}$ takes the value $\tilde{\xi}=16/27$. We applied this finding to the slowly rotating black hole candidate LMC X-3 for which $a_{\ast}=0.21^{+0.18}_{-0.22}$. Noting that $a^{\rm AS}_{\ast}$ is within the uncertainty range of $a_{\ast}$, is arguable that LMC X-3 could be understood as a RGI-Schwarzschild black hole with $\tilde{\xi}\simeq 16/27$. This conclusion should be taken with caution as it rests on the assumption that the relativistic thin disk model is appropriate to describe the accretion disk in LMC X-3, whereas the observational data show that this model is an accurate one at low luminosities but is inappropriate at high luminosities, a regime for which a slim accretion disk provides a better description \cite{r67,r68}.

It is important to remark, once again, that our results show that quantum gravity effects on the physics of black holes manifest themselves at distances even greater that the radius of the ISCO and are not restricted to the interior of the horizon or its immediate vicinity. We think that our results are a starting point to motivate the study of quantum gravity effects on more realistic black hole environments, such as accretion disks around a RGI-Kerr black hole, as an alternative or complementary path to confront the predictions of asymptotic safety with astrophysical observations. 

\section*{ACKNOWLEDGEMENTS}
We acknowledge financial support from COLCIENCIAS through the project with code 50754 associated to the {\it Convocatoria 757 para doctorados nacionales 2016}.

\end{document}